\documentclass[11pt,letterpaper]{article}
\usepackage{epsfig,rotating,setspace,latexsym,amsmath,epsf,amssymb,bm}
\usepackage{cite,graphicx,authblk,color,subfigure}

\title{Private Information Retrieval from Storage Constrained Databases   \\-- Coded Caching \textit{meets} PIR
\author{Maryam Abdul-Wahid \hspace{10pt} Firas Almoualem \hspace{10pt} Deepak Kumar \hspace{10pt} Ravi Tandon}
\affil{Department of Electrical and Computer Engineering\\
University of Arizona, Tucson, AZ, USA.\\
E-mail: {\{mabdulwahid, fmalem, deepakkumar, tandonr\}}@email.arizona.edu}}

\newtheorem{theorem}{Theorem}
\newtheorem{remark}{Remark}
\newtheorem{lemma}{Lemma}

\newtheorem{example}{Example}

\begin{document}
\maketitle
\newcommand\blfootnote[1]{%
  \begingroup
  \renewcommand\thefootnote{}\footnote{#1}%
  \addtocounter{footnote}{-1}%
  \endgroup
}

\blfootnote{This work was supported by the NSF Grant  CAREER-1651492.}

\thispagestyle{empty}
\vspace{-1.5cm}

\begin{abstract}
Private information retrieval (PIR) allows a user to retrieve a desired message out of $K$ possible messages from $N$ databases without revealing the identity of the desired message. There has been significant recent 
progress on understanding fundamental information theoretic limits of PIR, and in particular the download cost of PIR for several variations.  Majority of existing works however, assume the presence of replicated databases, each storing all the $K$ messages. In this work, we consider the problem of PIR from \textit{storage constrained} databases. Each database has a storage capacity of $\mu KL$ bits, where $K$ is the number of messages, $L$ is the size of each message in bits, and $\mu \in [1/N, 1]$ is the normalized storage.  

 In the storage constrained PIR problem, there are two key design questions: a) how to store content across each database under storage constraints; and b) construction of schemes that allow efficient PIR through storage constrained databases. The main contribution of this work is a general achievable scheme for PIR from storage constrained databases for any value of storage. In particular, for any $(N,K)$, with normalized storage $\mu= t/N$, where the parameter $t$ can take integer values $t \in \{1, 2, \ldots, N\}$, we show that our proposed PIR scheme achieves a download cost of $\left(1+ \frac{1}{t}+ \frac{1}{t^{2}}+ \cdots + \frac{1}{t^{K-1}}\right)$. The extreme case when $\mu=1$ (i.e., $t=N$) corresponds to the setting of replicated databases with full storage. For this extremal setting, our scheme recovers the information-theoretically optimal download cost characterized by Sun and Jafar as $\left(1+ \frac{1}{N}+ \cdots + \frac{1}{N^{K-1}}\right)$. For the other extreme, when $\mu= 1/N$ (i.e., $t=1$), the proposed scheme achieves a download cost of $K$. The most interesting aspect of the result is that for intermediate values of storage, i.e., $1/N < \mu <1$, the proposed scheme \textit{can strictly outperform} memory-sharing between extreme values of storage.

\end{abstract}

\vspace{-0.7cm}
\section{Introduction}
\label{sec:introduction}
Within the past few decades, there has been a surge in research towards solving various problems related to private information retrieval. The goal behind the private information retrieval (PIR) problem is to determine the most efficient solution that allows a user to retrieve a certain message from a set of distributed databases - each contains multiple messages - without any of those databases determining which message has been requested. Since the introduction of PIR in \cite{ChorAndCompany}, this problem has received significant attention in the computer science community \cite{TrinabhGupta, Demmler, CachinAndCompany, Yekhanin} and PIR protocols have found use in information-theoretic security, oblivious transfer protocols, locally decodable codes and numerous other areas.
The classical PIR problem involves $N$ non-colluding databases, where each stores $K$ messages. A user requests a message by generating a query to each database. The databases each respond to the user with an answer. Then the user must be able to correctly obtain the desired message from all $N$ answers. To ensure privacy, every query and every answer is independent of the requested message.  Based on the Shannon theoretic formulation, the rate of the private information retrieval problem is set as the number of desired information bits per number of downloaded bits. The information theory capacity, $C$, is then the maximum PIR rate possible. Previous works observe the PIR rate with full storage among the databases, i.e. each database stores every message.  Under this assumption, one of the first achievable PIR rate was found by Shah, Rashmi and Ramchandran \cite{NiharPIR} to equal $1-\frac{1}{N}$.  

In a very interesting recent work \cite{SunAndJaffar1}, Sun and Jafar characterized the exact information-theoretic capacity (or the inverse of download cost) of the $(N,K)$ PIR problem as $(1 + 1/N + \ldots 1/N^{K-1})^{-1}$, improving upon the previous best known achievable rate for the PIR problem \cite{NiharPIR}. Since the appearance of \cite{SunAndJaffar1}, significant progress has been made on a variety of variations of the basic PIR problem. We briefly describe some of these advances next. The case of $T$-colluding PIR (or TPIR in short) was investigated in \cite{SunAndJaffar2}, where any $T$ databases out of $N$ are able to collude, i.e., they can share the queries. Robust PIR, in which  any $N$ out of $M$ databases (with $N\leq M$) fail to respond was also investigated in \cite{SunAndJaffar2}, for which the capacity is found to be the same as that of TPIR. In a recent work, \cite{BanawanAndUlukus2} characterized the  capacity of PIR with byzantine databases (or BPIR), i.e., a scenario in which any $L$ out of $N$ databases are adversarial (i.e. they can respond with incorrect bits after receiving the query). The above previous works assumed the presence of replicated databases, i.e., each database stores all the $K$ messages. The capacity of PIR with databases storing MDS coded messages was considered in \cite{Salim} and the capacity was subsequently characterized by Banawan and Ulukus in \cite{BanawanAndUlukus}. This setting was further investigated for the scenario where any $T$ out of $N$ databases can collude, an aspect termed MDS-TPIR \cite{FreijGnilkeHollantiKarpuk, SunAndJafar3} although its capacity remains open for general set of parameters. The problem of symmetric  PIR (SPIR) was studied in \cite{SunAndJafar4}. In this setting, privacy is enforced in both directions: i.e.,  user must be able to retrieve the message of interest  privately while at the same time the databases must avoid any information leakage to the user about the remaining messages. The exact capacities for this symmetric PIR problem both  for non-coded (SPIR) and MDS-coded (MDS-SPIR) messages were characterized in \cite{WangAndSkoglund, SunAndJafar4}. The case of multi-message PIR (MPIR) was investigated in \cite{BanawanAndUlukus3, ZhangAndGennian}, in which the user wants to privately retrieve $P\geq 1$ out of $K$ messages. The capacity of cache constrained PIR (in which the user has a local cache of limited storage) was recent characterized in \cite{TandonCachePIR}, and it was shown that memory sharing based PIR scheme is information theoretically optimal (also see recent works \cite{Jafar-NewCache, Ulukus-NewCache, Kadhe-NewCache} on other variations of the cache aided PIR problem). Majority of above works however, assume the presence of replicated databases, each storing all the $K$ messages. Indeed, exceptions to this statement include the work on the case when database store MDS coded data, and the databases must also satisfy the $k$-out-of-$N$ recovarability constraint. Furthermore, \cite{SunJafarlimited} also investigated the problem of limited storage PIR for the special case of $K=2$ messages and $N=2$ databases. They present interesting lower and upper bounds on the capacity for this special case, and show the optimality of the proposed scheme for the case of linear schemes. However, at this point, generalization of the scheme for $(N,K)=(2,2)$ to arbitrary $(N,K)$ remains elusive.

\textbf{Summary of Contribution and Insights--} In this work, we consider the problem of PIR from \textit{storage constrained} databases. Each database has a storage capacity of $\mu KL$ bits, where $K$ is the number of messages, $L$ is the size of each message in bits, and $\mu \in [1/N, 1]$ is the normalized storage.  
In the storage constrained PIR problem, there are two key design questions: a) how to place content across each database under storage constraints; and b) construction of schemes that allow efficient PIR through storage constrained databases. The main contribution of this work is an achievable scheme for PIR from storage constrained databases for any arbitrary $(N,K)$, and any value of storage. In particular, for any $(N,K)$, with normalized storage $\mu= t/N$, where the parameter $t$ can take integer values $t \in \{1, 2, \ldots, N\}$, we show that our proposed PIR scheme achieves a download cost of $(1+ \frac{1}{t}+ \frac{1}{t^{2}}+ \cdots + \frac{1}{t^{K-1}})$. 

There are two main ingredients in our storage constrained PIR scheme which we briefly explain next. The first ingredient is the storage strategy across databases, which is inspired by the work of Maddah-Ali and Niesen \cite{Maddah} on the fundamental limits of caching. In particular, we split each message into a number of sub-messages (also known as sub-packets in the caching literature), and index each sub-message by a sub-set of the $N$ databases, which end up storing the sub-message. The amount of  per-message sub-packetization is carefully chosen to satisfy the storage constraint at each database. As it turns out, this storage strategy naturally helps in the design of the second main ingredient, i.e., the design of efficient PIR by utilizing the limited storage.  To this end, we tailor the key components of (full storage) PIR scheme of Sun and Jafar to the case of limited storage setting, by enforcing symmetry across messages, and exploiting side information from other databases. The main differences from that of Sun and Jafar are two fold: a) from a privacy perspective, enforcement of message symmetry is necessary only across sub-packets for each message that are stored at a database;  b) on the other hand, side information can only be partially exploited depending on the contents shared across databases, which in turn depends on the amount of storage. 
The extreme case when $\mu=1$ (i.e., $t=N$) corresponds to the setting of replicated databases with full storage. For this extremal setting, our scheme recovers the information-theoretically optimal download cost characterized by Sun and Jafar as $(1+ \frac{1}{N}+ \cdots + \frac{1}{N^{K-1}})$. For the other extreme, when $\mu= 1/N$ (i.e., $t=1$), the proposed scheme achieves a download cost of $K$, which is information theoretically optimal. The most interesting aspect of the result is that for intermediate values of storage, i.e., $1/N < \mu <1$, the proposed scheme \textit{strictly outperforms} memory-sharing.

\section{Storage Constrained Information-Theoretic PIR}
\label{sec:problem}

Consider a PIR problem where there are $N$ non-colluding databases and $K$ independent messages $W_1,W_2,\ldots,W_K$, where each message is of size $L$ bits, i.e., 
\begin{equation}
H(W_1)=H(W_2)=\ldots=H(W_K)=L
\end{equation}
We assume that each database has a storage capacity of $\mu KL$ bits. If we denote $Z_1, Z_2, \ldots, Z_N$ as the contents stored across the databases, then we have the following storage constraint for each database:
\begin{equation}
H(Z_1)=H(Z_2)=\ldots=H(Z_N)=\mu K L
\end{equation}
We allow the user to design what contents can be stored at each database subject to the storage constraint. Furthermore, we assume that the storage strategy employed by the user is completely public i.e., each database knows which contents are stored at all the other databases. The normalized storage parameter $\mu$ can take values in the range $1/N \leq \mu \leq 1$. The case when $\mu=1$ is the setting of replicated databases, with each database storing all the $K$ messages. The lower bound $\mu \geq 1/N$ is in fact a necessary condition for reliable decoding.

\begin{figure}[t]
	\begin{center}
		\includegraphics[width=0.7\columnwidth]{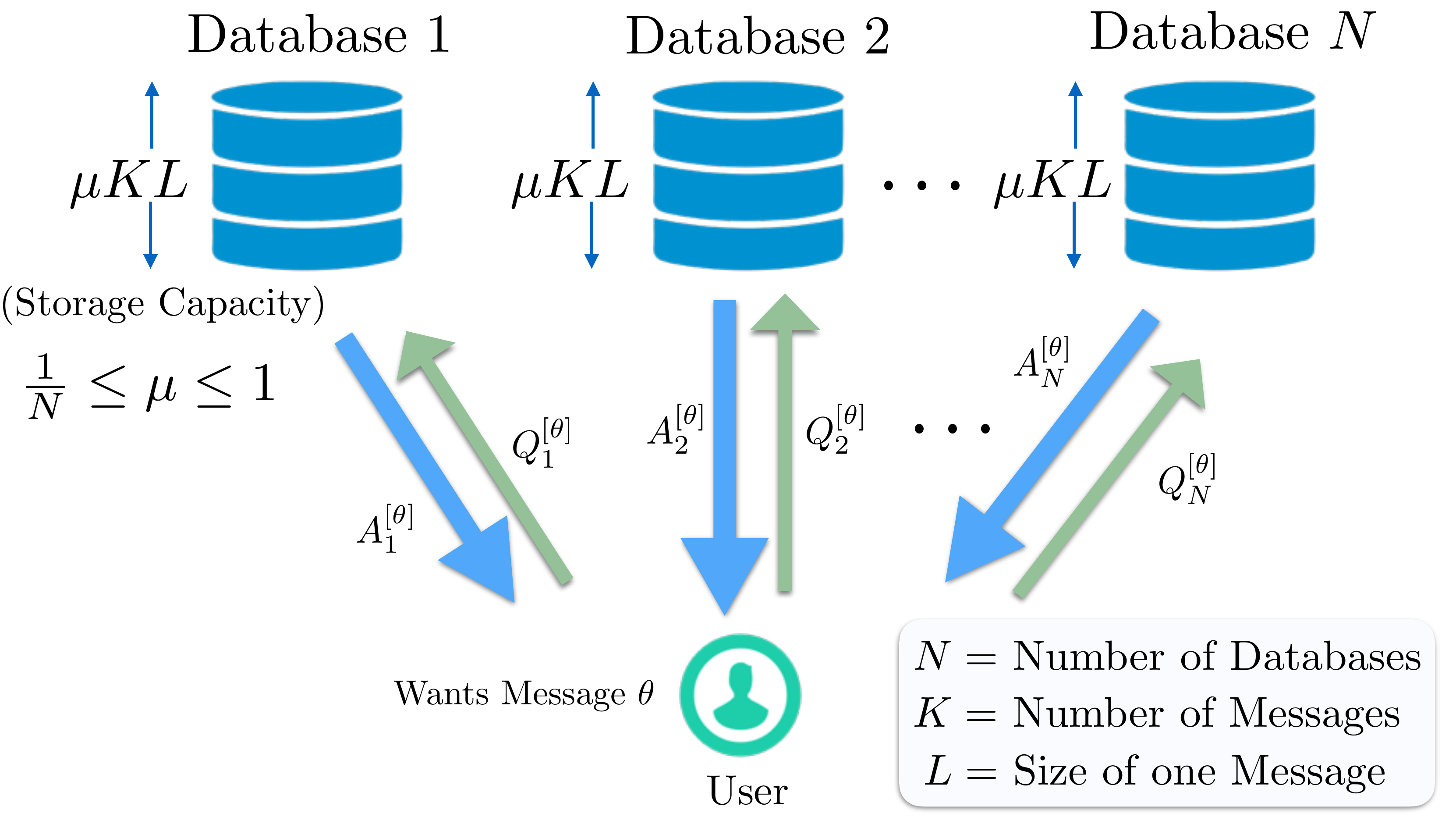}
		\caption{Storage Constrained Private Information Retrieval. \label{Figmodel}}
		\vspace{-15pt}
	\end{center}
\end{figure}

To request a message, a user privately selects a number $\theta$ between $1$ and $K$ to correspond with the desired message $W_{\theta}$. Then the user generates $N$ queries $Q_1^{[\theta]}, Q_2^{[\theta]},\ldots, Q_N^{[\theta]}$ that will be sent to the $N$ databases. The superscript $\theta$ indicates the desired message, which can be substituted with any of the $K$ messages. Privacy must be ensured from the moment the user decides on the message being requested.  Therefore, the queries must be independent of the messages.
\begin{equation} \forall k\in[K],\, I(W_1,W_2,\ldots,W_K;Q_1^{[k]},Q_2^{[k]},\ldots,Q_N^{[k]})=0
\end{equation}
The user requests message $W_k$ by sending a query $Q_n^{[k]}$ to the $n$-th database, which then generates and returns an answer $A_n^{[k]}$ back to the user. The answer is a function of the corresponding query and the data stored in the $n$-th database.
\begin{equation}
\forall k\in[K],\forall n\in[N],\:H(A_n^{[k]}|Q_n^{[k]},Z_n)=0
\end{equation}

From all of the answers from each database, the user must be able to correctly decode the desired message $W_k$ with a small probability of error. The correctness constraint is as follows
\begin{equation}
 H(W_k|A_1^{[k]},\ldots,A_N^{[k]},Q_1^{[k]},\ldots,Q_N^{[k]})=o(L)
\end{equation}
where $o(L)$ represents a function of $L$ such that $o(L)/L$ approaches $0$ as $L$ approaches infinity. 

In order to prevent the databases from learning which message has been requested, the following privacy constraint must be satisfied.
\begin{align}
(Q_n^{[i]},A_n^{[i]},W_1,\ldots,W_K, Z_1, \ldots, Z_n)
 \sim(Q_n^{[j]},A_n^{[j]},W_1,\ldots,W_K, Z_1, \ldots, Z_n),\forall i\neq j.
\end{align}

For a storage parameter $\mu$, we say that the pair $(D,L)$ is achievable if there exists a Storage Constrained PIR scheme with storage, querying, and decoding functions, which satisfy the storage, correctness and privacy constraints. The performance of a PIR scheme is characterized by the number of bits of desired information ($L$) per downloaded bit. In particular, if $D$ is the total number of downloaded bits, and $L$ is the size of the desired message, then the normalized downloaded cost is $D/L$. In other words, the PIR rate is $L/D$. 

The goal is to characterize the optimal normalized download cost as a function of the per-database storage parameter $\mu$:
\begin{align}
D^{*}(\mu)= \text{min }\{D/L: (D,L) \text{ is achievable}\}.
\end{align}
The storage-constrained capacity of PIR is the inverse of the normalized download cost 
\begin{align}
C^{*}(\mu)= \text{max }\{L/D: (D,L) \text{ is achievable}\}.
\end{align}

We first state and prove the following Lemma which shows that the optimal download cost $D^{*}(\mu)$ (or the inverse of capacity $1/C^{*}(\mu)$) is a convex function of the normalized storage $\mu$. 

\begin{lemma}\label{Lemma1}
The optimal download cost $D^{*}(\mu)$ is a convex function of $\mu$. In other words, for any $(\mu_1, \mu_2)$, and $\alpha \in [0,1]$, the optimal download cost satisfies
\begin{align}
D^{*}(\alpha \mu_1 + (1-\alpha)\mu_2) \leq \alpha D^{*}(\mu_1) + (1-\alpha)D^{*}(\mu_2).
\end{align}
\end{lemma}

\textit{Proof of Lemma \ref{Lemma1}--} Let us consider two storage parameters $\mu_1$, and $\mu_2$, with optimal download costs $D^{*}(\mu_1)$, and $D^{*}(\mu_2)$ respectively using two storage constrained PIR schemes, say Scheme $1$ and Scheme $2$. Let us now consider a new storage point $\mu= \alpha \mu_1 + (1-\alpha) \mu_2$, for which we can construct a PIR scheme as follows: we take each message $W_i$ and divide it into two independent parts $W_i = \left(W_i^{(1)}, W_i^{(2)}\right)$, where $W_{i}^{(1)}$ is of size $L_{1}=\alpha L$ bits, and $W_{i}^{(2)}$ is of size $L_{2}=(1-\alpha)L$ bits. The total size of each message is hence $L$. For each sub-message, the databases utilize the storage and PIR schemes $1$, and $2$ respectively. In particular, Scheme $1$ requires a storage of $\mu_1 \alpha KL$ bits, and Scheme $2$ requires a 
storage of $\mu_2 (1-\alpha)KL$ bits. Furthermore, PIR from Scheme $1$ requires a download of $D^{*}(\mu_1)L_1= \alpha D^{*}(\mu_1)L$ bits and scheme $2$ requires a download of $(1-\alpha)D^{*}(\mu_2)L$ bits. 
Hence, the total downloaded data is therefore $(\alpha D^{*}(\mu_1) + (1-\alpha)D^{*}(\mu_2))L$, and hence the normalized download cost for this memory-sharing scheme is $(\alpha D^{*}(\mu_1) + (1-\alpha)D^{*}(\mu_2))$. Furthermore, the total storage used by each database is $S= (\alpha\mu_1 + (1-\alpha)\mu_2)K L= \mu KL $, i.e., the normalized storage parameter is $\mu= \alpha \mu_1 + (1-\alpha) \mu_2$. 
Since $D^{*}(\alpha \mu_1 + (1-\alpha)\mu_2)$ by definition is optimal download cost for normalized storage $(\alpha \mu_1 + (1-\alpha)\mu_2)$, it must be upper bounded by the download cost of the memory sharing scheme. Hence, the following inequality follows
\begin{align}
D^{*}(\alpha \mu_1 + (1-\alpha)\mu_2) \leq \alpha D^{*}(\mu_1) + (1-\alpha)D^{*}(\mu_2),
\end{align}
which proves the convexity of $D^{*}(\mu)$. 

\section{Main Result and Discussions}\label{sec:results}
We next present the main result of this paper in the following Theorem.
\begin{theorem} \label{theorem1}
For the storage constrained PIR problem with $N$ Databases, $K$ messages, (of size $L$ bits each), and a per database storage constraint of $\mu KL$ bits, the lower convex hull of the following $(\mu, D(\mu))$ pairs is achievable. 
	\begin{align}
	\label{eq:theorem1}
	{(\mu,\ D(\mu))}=\left(\frac{t}{N},  \ 1+\frac{1}{t}+\cdots +\frac{1}{t^{K-1}}\right),
	\end{align} 
for $t=1, 2, \ldots, N$. 
\end{theorem}

\begin{figure}[t]
  \begin{center}
  \includegraphics[width=0.75\columnwidth]{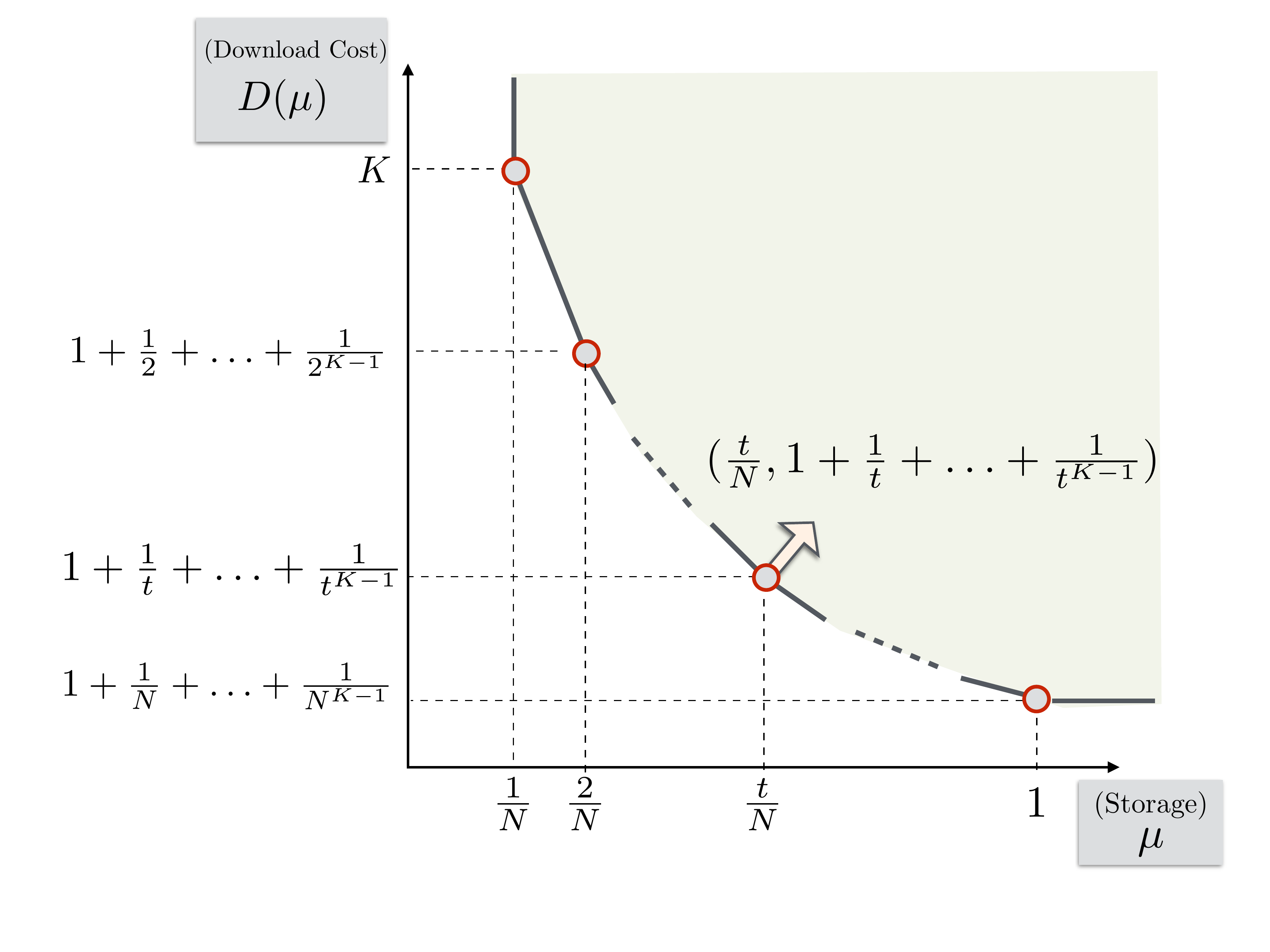}
\caption{The tradeoff between storage and download cost for PIR. \label{fig1}}
\vspace{-10pt}
  \end{center}
\end{figure}

\begin{remark}\label{remark1}
The general tradeoff resulting from Theorem \ref{theorem1} is illustrated in Fig. \ref{fig1}. The smallest value of $\mu=1/N$ corresponds to the parameter $t=1$, for which the download cost is maximal and is equal to $K$. The other extreme value of storage is $\mu=1$, corresponding to  $t=N$, i.e., the setting of full storage in which every database can store all the messages. For this case, the optimal download cost was characterized by Sun and Jafar \cite{SunAndJaffar1} as $(1+\frac{1}{N}+\frac{1}{N^2}+.....+\frac{1}{N^{K-1}})$, where N is number of databases. 
\end{remark}

\noindent The proof of  Theorem \ref{theorem1} has two main parts: a) the storage design (i.e., how to store content across $N$ databases) subject to storage constraints; and b) the design of the PIR scheme from storage constrained databases. We next describe our storage scheme while satisfying the constraint that each database has a storage capability of at most $\mu KL$ bits. 

\textbf{Storage Scheme for $\mu = t/N$:} 
For a fixed parameter $t\in [1,\ldots, N]$, we take each message $W_i$ and sub-divide it into $\binom{N}{t}$ sub-messages. In particular, each sub-message is indexed by a subset of databases of size $t$. For instance, if $t=2$, and $N=3$, then each message $W_i$ will be sub-divided into $\binom{3}{2}= 3$ sub-messages as $W_{i}=(W_{i, \{1,2\}}, W_{i, \{2,3\}}, W_{i, \{1,3\}})$. Furthermore, we assume that each sub-message is of size $t^{K}$ bits. Hence the total size of each message, i.e., $L$ is given as $L= \binom{N}{t}t^{K}$. Using this message splitting scheme, we propose the storage scheme as follows: for every message, each database stores all sub-messages which contain its index. For instance, for the $t=2$, $N=3$ databases, and $K=2$ messages (say $A$, and $B$), we split the messages as $A=(A_{12}, A_{23}, A_{13})$, and $B=(B_{12}, B_{23}, B_{13})$, and hence, the storage strategy is as follows:
\begin{itemize}
\item DB$_1$ stores $A_{12}, A_{13}, B_{12}, B_{13}$
\item DB$_2$ stores $A_{12}, A_{23}, B_{12}, B_{23}$
\item DB$_3$ stores $A_{13}, A_{23}, B_{13}, B_{23}$
\end{itemize}
We next verify that the above scheme satisfies the storage constraint. To this end, we note that for every message, each database stores $\binom{N-1}{t-1}$ sub-messages (this corresponds to the number of sub-sets of databases of size $t$ in which the given database is present). 
Hence, the total storage necessary for any database is given as:
\begin{align}
K\times \binom{N-1}{t-1}\times t^{K}&= \frac{t}{N}\times K\times N\times \binom{N-1}{t-1}\times t^{K-1}\nonumber\\
&= \frac{t}{N}\times K\times \left(\binom{N}{t}t^{K}\right)\nonumber\\
&= \frac{t}{N}\times K\times L= \mu KL\nonumber
\end{align}
This shows that the proposed scheme satisfies the storage constraints for every database.  

Before presenting the  proof of the general PIR scheme for any $(N,K, \mu)$, we first provide some representative examples which highlight the key new ideas and novel aspects that are necessary for the general scheme.  Our storage constrained PIR scheme is similar in spirit to the scheme of Sun and Jafar in the sense that we also enforce symmetry across databases and message, and the exploitation of undesired side information. The main difference from Sun and Jafar's scheme is that in storage constrained scenario, each database may not have access to all of the side information retrieved from other databases. Hence, only those side information bits can be exploited which are shared between databases. These similarities and differences will become more clear in following examples. 

\vspace{5pt}
\hrule
\vspace{5pt}

	\begin{figure}[t]
  \begin{center}
  \includegraphics[width=0.65\columnwidth]{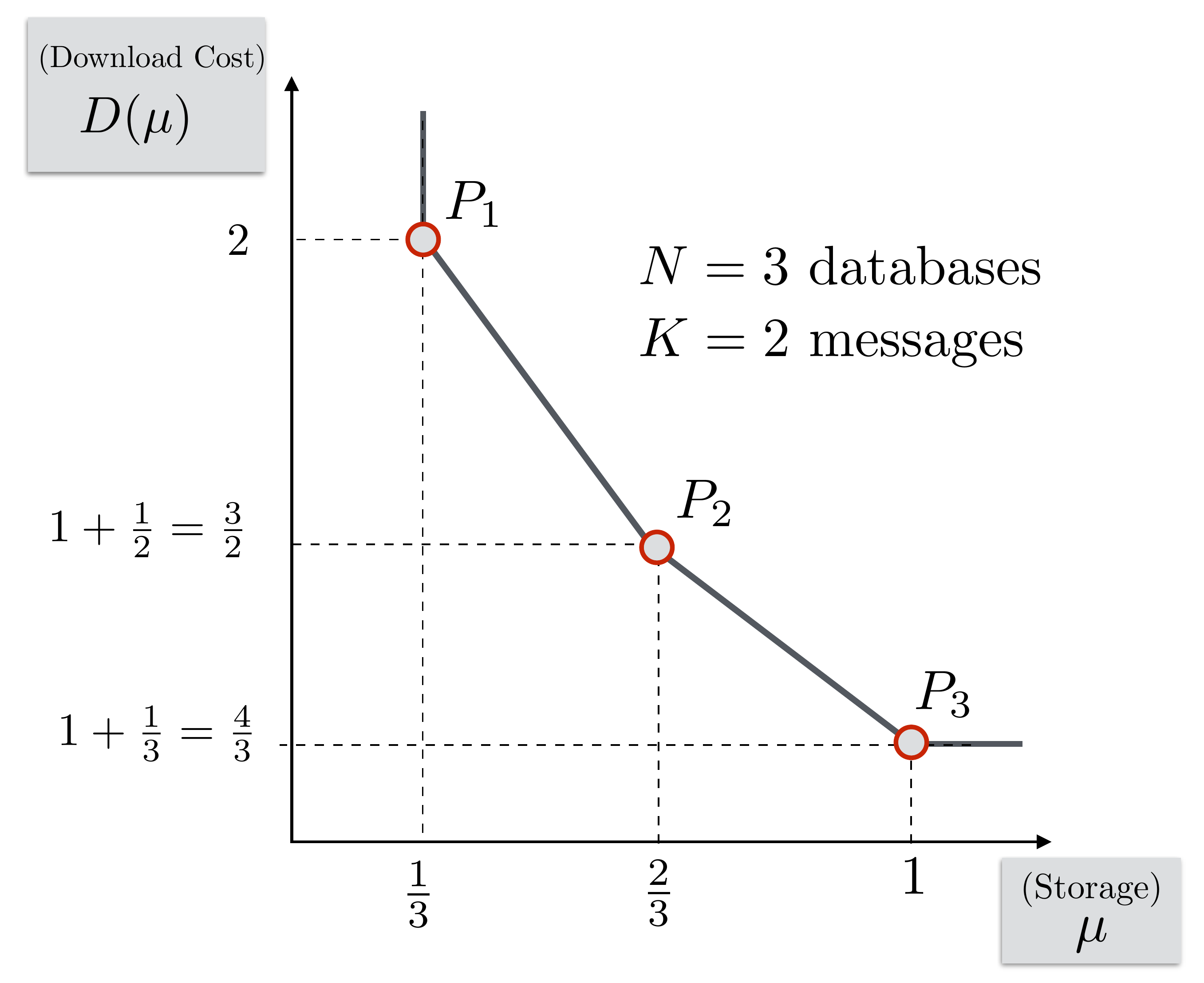}
\vspace{-10pt}
\caption{Storage vs. download cost for $(N,K)=(3,2)$. \label{fig2}}
\vspace{-10pt}
  \end{center}
\end{figure}

\begin{example}[$N=3$ databases, $K=2$ messages]\normalfont\label{ex2}
	In this example, we have 3 databases \break $(DB_1,\ DB_2,\ DB_3)$, and $K=2$ messages which we denote as $A$ and $B$. As shown in Fig. \ref{fig2}, the tradeoff has three critical points (labeled as $P_1$, $P_2$ and $P_3$). 
The point $P_1$ corresponds to $\mu=1/3$, for which the PIR scheme is trivial: download all $K=2$ messages. On the other extreme, point $P_3$ corresponds to $\mu=1$, for which the optimal PIR scheme is due to Sun and Jafar, with a download cost of $4/3$. For the sake of illustrative purposes, let us first revisit this scheme: each message is divided into $N^{K}= 3^{2}=9$ bits (i.e., the messages $A$ and $B$ are represented by $A=(a_1, a_2, \ldots, a_9)$, and $B=(b_1, b_2, \ldots, b_9)$. Suppose that the user wants to retrieve message $A$ privately. Then, the scheme of Sun-Jafar works as shown in Table \ref{Sun-Jafar}. The user downloads single bits from each database for each of the two messages. In the second stage, the user downloads the XOR's of the bits of both messages and exploits side-information (i.e., the undesired bits of message $B$) downloaded from the remaining $(N-1)=2$ databases. 
\begin{table} [!h]
		\renewcommand{\arraystretch} {1.3}
		\caption{Sun-Jafar scheme for $(N,K)=(3,2)$}
		\label{Sun-Jafar}
		\centering 
		\begin{tabular}{|c|c|c|}
			\hline
			DB1 &DB2 & DB3 \\
			\hline
			$a_1  \  \ b_1$ \ & $a_2 \ \ b_2$ \ & $a_3 \  \ b_3$\\
			\hline
			$a_4+ b_2$ \ & $a_6+ b_1$ \ & $a_8 + b_1$\\
			$a_5+ b_3$ \ & $a_7 + b_3$ \ & $a_9 + b_2$\\
		\hline
		\end{tabular}
	\end{table}
The user downloaded a total of $12$ bits, out of which the desired bits are $9$ (corresponding to message $A$). Hence, the download cost of this scheme is $D(1)= 12/9 = 4/3$.

\noindent \underline{Storage Constrained PIR scheme for $\mu=2/3$ (point $P_2$):}  We next present the scheme for $\mu=\frac{2}{3}$, i.e., $t=2$. Since $t=2$, the total size of each message is  chosen as $L=2^{2}\binom{3}{2}=12$ bits. Total storage used by each database is given as $S=\mu KL=\frac{2}{3}\times2 \times12=16$ bits, and the data stored across each database is shown in Table \ref{table32}.

	\begin{table} [!h]
		\renewcommand{\arraystretch} {1.3}
		\caption{Storage for $(N,K)=(3,2)$ and $\mu=2/3$ ($t=2$) }
		\label{table32}
		\centering 
		\begin{tabular}{|c|c|c|}
			\hline
			$DB_1$ & $DB_2$ & $DB_3$ \\
			\hline
			 $a^1_{12}\ b^1_{12}$   &$a^1_{12}\ b^1_{12}$  &$a^1_{13}\ b^1_{13}$   \\
			 $a^1_{13}\ b^1_{13}$ &  $a^1_{23}\ b^1_{23}$ & $a^1_{23}\ b^1_{23}$   \\
			 $a^2_{12}\ b^2_{12}$   &$a^2_{12}\ b^2_{12}$  &$a^2_{13}\ b^2_{13}$   \\
			$a^2_{13}\ b^2_{13}$ &  $a^2_{23}\ b^2_{23}$ & $a^2_{23}\ b^2_{23}$  \\
			 $a^3_{12}\ b^3_{12}$   &$a^3_{12}\ b^3_{12}$  &$a^3_{13}\ b^3_{13}$   \\
			$a^3_{13}\ b^3_{13}$ &  $a^3_{23}\ b^3_{23}$ & $a^3_{23}\ b^3_{23}$  \\
			$a^4_{12}\ b^4_{12}$   &$a^4_{12}\ b^4_{12}$  &$a^4_{13}\ b^4_{13}$ \\
			$a^4_{13}\ b^4_{13}$ &  $a^4_{23}\ b^4_{23}$ & $a^4_{23}\ b^4_{23}$ \\			
			\hline
		\end{tabular}
	\end{table}
\noindent Let us now assume that we want to retrieve message $A$ privately. We start downloading single bits 
$(a^1_{12},\ a^1_{13})$ from $DB_1$ and for message symmetry $(b^1_{12},\ b^1_{13})$ is also downloaded from $DB_1$. Similarly, we download $(a^2_{12},\ a^1_{23}),(b^2_{12},\ b^1_{23})$ from $DB_2$, and $(a^2_{13},\ a^2_{23}),\ (b^2_{13},\ b^2_{23})$ from $DB_3$ is downloaded (see Table ~\ref{table}). 
	\begin{table} [!htbp]
		\renewcommand{\arraystretch} {1.3}
		\caption{Storage Constrained PIR Scheme $(\mu=2/3)$}
		\label{table}
		\centering 
		\begin{tabular}{|c|c|c|}
			\hline
			DB1 &DB2 & DB3 \\
			\hline
			$a^1_{12}\  \ b^1_{12}$   &$a^2_{12}\ \ b^2_{12}$  &$a^2_{13}\ \ b^2_{13}$   \\
			$a^1_{13}\  \ b^1_{13}$ &  $a^1_{23}\ \ b^1_{23}$ & $a^2_{23}\ \ b^2_{23}$ \\
			\hline
			$a^3_{12}+b^2_{12}$\  & $a^4_{12}+b^1_{12}$\  &$a^4_{13}+b^1_{13}$\\\
			$a^3_{13}+b^2_{13}$ & $a^3_{23}+b^1_{23}$ &  $a^4_{23}+b^1_{23}$\\
			\hline
		\end{tabular}
	\end{table}
We next move to the second stage of the scheme, in which the user exploits the side information (or undesired bits of message $B$). However, the main difference from full storage is that not all side-information can be exploited from every database due to the storage constraint. In particular, we can notice that from the perspective of database $1$, only those bits of the undesired message $B$ can be leveraged as side information that are stored at $DB_1$, i.e., the bits which have $DB_1$ as one of the index in the subset of $t$ databases that have stored that bit. For this example, only the bits $b^{2}_{12}$ and $b^{2}_{13}$ can be leveraged as side information through $DB_1$. 

From Table \ref{table}, we can count the number of useful bits downloaded as  $12$ and total number of bits downloaded $18$. This leads to the download cost of $\frac{18}{12}= \frac{3}{2}$ which matches the result of Theorem \ref{theorem1}. 	
\end{example}

\vspace{5pt}
\hrule
\vspace{5pt}

\begin{example}[$N=3$ databases, $K=3$ messages]\normalfont\label{ex3}
	In this example, we have 3 databases \break $(DB_1,\ DB_2,\ DB_3)$, and $K=3$ messages which we denote as $A, B$ and $C$. As shown in Fig. \ref{fig3}, the tradeoff has three critical points (labeled as $P_1$, $P_2$ and $P_3$). As the point $P_1$ is trivial and $P_3$ follows from the work of Sun and Jafar, we explain the achievability of the point $P_2$. For this point, we have $t=2$, and hence each message is of size $L=t^{K}\binom{N}{t}= 2^{3}\times 3= 24$ bits. In particular, we split each message into $\binom{3}{2}=3$ sub-messages, and each sub-message is of size $2^{3}=8$ bits as follows:

\begin{itemize}
\item $A=(a^{1}_{12},\ldots, a^{8}_{12}, \ \ a^{1}_{23},\ldots, a^{8}_{23}, \ \ a^{1}_{13},\ldots, a^{8}_{13})$
\item $B=(b^{1}_{12},\ldots, b^{8}_{12}, \ \ b^{1}_{23},\ldots, b^{8}_{23}, \ \ b^{1}_{13},\ldots, b^{8}_{13})$
\item $C=(c^{1}_{12},\ldots, c^{8}_{12}, \ \ c^{1}_{23},\ldots, c^{8}_{23}, \ \ c^{1}_{13},\ldots, c^{8}_{13})$
\end{itemize}	
Subsequently,  $DB_i$ stores those bits (of each message) whose index contains $i$. Hence, the  total storage required per DB is $3\times 2\times 8 = 48$ bits. With the storage designed, the PIR scheme works as follows in three stages:

\noindent \textbf{Stage $1$}: In this stage, we download single bits from the messages. From $DB_1$, we download ($a^1_{12},\ a^1_{13}$), then we download $(b^1_{12} ,\  b^1_{13})$, and $(c^1_{12},\  c^1_{13})$ to maintain message symmetry. Similarly, $(a^2_{12},\ a^1_{23},\ b^2_{12},\ b^1_{23},\ c^2_{12},\  c^1_{23})$ and $(a^2_{13},\ a^2_{23},\ b^2_{13} ,\  b^2_{23},\ c^2_{13},\   c^2_{23})$ are downloaded from $DB_2$ and $DB_3$ respectively.

\noindent \textbf{Stage $2$}: In the second stage, we download pairs of bits with the goal of maximally utilizing the side information (of undesired messages) in the previous stage.  Again, the key aspect to note here is that when downloading bits from $DB_1$'s, we can only utilize those bits as side information which are stored at it. For this example, we can utilize $b^{2}_{12}, c^{2}_{12}, b^{2}_{13}, c^{2}_{13}$ since these bits were downloaded from $DB_2$ and $DB_3$ in Stage $1$, and are also stored at $DB_1$. As shown in  Table \ref{table33}, we download $(a^3_{12}+b^2_{12},\ a^4_{12}+c^2_{12},\ a^5_{13}+b^2_{13},\ a^6_{13}+c^2_{13})$ from $DB_1$. In order to maintain message symmetry, we also download $(b^3_{12}+c^3_{12},\ b^3_{13}+c^3_{13})$ from $DB_1$. We follow a similar process across $DB_2$ and $DB_3$. 

	\begin{figure}[t]
  \begin{center}
  \includegraphics[width=0.65\columnwidth]{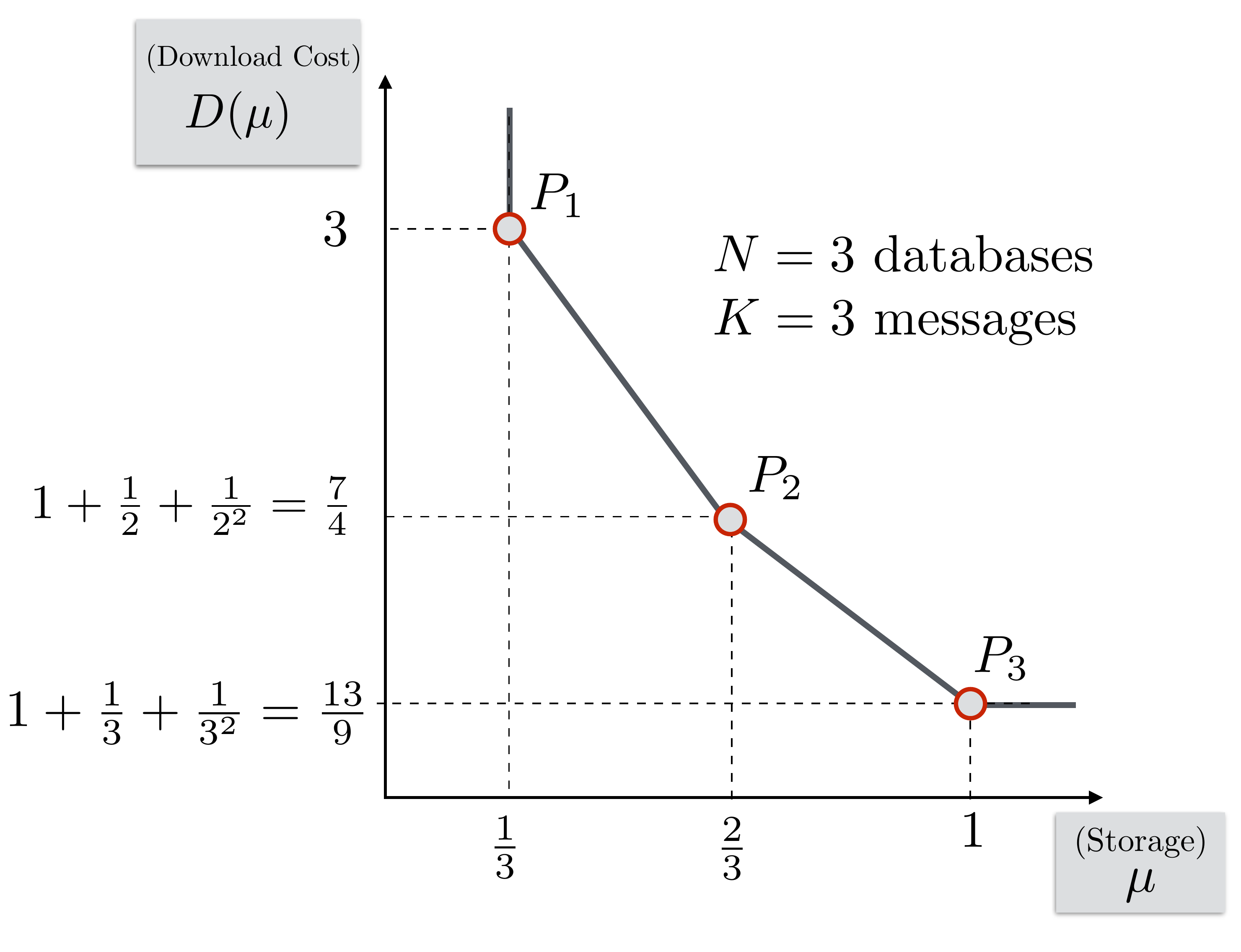}
\vspace{-10pt}
\caption{Storage vs. download cost for $(N,K)=(3,3)$. \label{fig3}}
\vspace{-0pt}
  \end{center}
\end{figure}
%\vspace{-20pt}
	\begin{table} [!h]
		\hrule \vspace{4pt}
		\renewcommand{\arraystretch} {1.3}
		\caption{\hspace{-1pt}Storage Constrained PIR: $(N,K)=(3,3), \mu= 2/3$}\vspace{4pt}
		\label{table33}
		\centering 
		\begin{tabular}{|c|c|c|}
			\hline
			DB1 &DB2 & DB3 \\
			\hline
			$a^1_{12}\ b^1_{12}\ c^1_{12}$   &$a^2_{12}\ b^2_{12}\ c^2_{12}$  &$a^2_{13}\ b^2_{13}\ c^2_{13}$   \\
			$a^1_{13}\ b^1_{13}\ c^1_{13}$ &  $a^1_{23}\ b^1_{23}\ c^1_{23}$ & $a^2_{23}\ b^2_{23} \ c^2_{23}$ \\
			\hline
			$a^3_{12}+b^2_{12}$\  & $a^5_{12}+b^1_{12}$\  &$a^5_{13}+b^1_{13}$\\
			$a^3_{13}+c^2_{13}$ & $a^3_{23}+c^1_{12}$ &  $a^5_{23}+c^1_{13}$\\
			$a^4_{12}+b^2_{13}$\  & $a^6_{12}+b^2_{23}$\  &$a^6_{13}+b^1_{23}$\\
			$a^4_{13}+c^2_{13}$ & $a^4_{23}+c^2_{23}$ &  $a^6_{23}+c^1_{23}$\\
			$b^3_{12}+c^3_{12}$ & $b^4_{12}+c^4_{12}$ &  $b^4_{13}+c^4_{13}$\\
			$b^3_{13}+c^3_{13}$ & $b^3_{23}+c^3_{23}$ &  $b^4_{23}+c^4_{23}$\\
			\hline
			$a^7_{12}+b^4_{12}+c^4_{12}$ &$a^8_{12}+b^3_{12}+c^3_{12}$ &  $a^8_{13}+b^4_{12}+c^4_{12}$\\
			$a^7_{13}+b^4_{13}+c^4_{13}$ &$a^7_{23}+b^4_{23}+c^4_{23}$ &  $a^8_{13}+b^3_{23}+c^3_{23}$\\
			\hline
		\end{tabular}
\vspace{7pt}\hrule 
	\end{table}

\noindent \textbf{Stage $3$}: In the final stage, we download triples of bits  (i.e., $a+b+c$'s) from each database. Following the same principle as before, we can observe that when downloading from $DB_1$, we can  leverage $(b^{4}_{12}+ c^{4}_{12})$ and $(b^{4}_{13}+ c^{4}_{13})$ as side information which was downloaded in Stage $2$, and is available at $DB_1$. As shown in Table \ref{table33}, we download $a^{7}_{12}+b^{4}_{12}+ c^{4}_{12}$ and $a^{7}_{13}+b^{4}_{13}+ c^{4}_{13}$ from $DB_1$. We follow a similar process for the other two databases. 
	
We can readily verify that from the data downloaded  from all three databases (which is $42$ bits), the user is able to correctly retrieve the message $A$, i.e., the user is able to decode all $24$ desired bits $(a^{1}_{12},\ldots, a^{8}_{12}, \ \ a^{1}_{23},\ldots, a^{8}_{23}, \ \ a^{1}_{13},\ldots, a^{8}_{13})$. Hence, the download cost of our scheme is $D(\frac{2}{3})= \frac{42}{24}=7/4$ which is equal to the expression in Theorem \ref{theorem1} and also shown in Fig. \ref{fig3} (point $P_2$). 	
	
\end{example}

\newpage
\section{Proof of Theorem 1 (Storage Constrained PIR Scheme)}
In this section, we present the proof of the general storage constrained PIR scheme for any $(N,K)$. We focus on the storage parameter $\mu=t/N$ for any $t\in [1:N]$.  In the general scheme, we follow the same philosophy introduced in the examples. Namely, the general scheme works in a sequence of $K$ stages, where in each stage, we download tuples of bits by exploiting side information obtained from the previous stage, maintain message symmetry, as well as symmetry across databases. Most importantly, the exploitation of side information is carefully designed to account for the limited storage capabilities of the databases. We next present the general scheme, and assume that we want to privately retrieve message $1$. 

\vspace{8pt}
\noindent \textbf{Stage $1$}: In the first stage, we start downloading single bits of each message from each database. Let us focus on a single database (say $DB_1$). From this database, we first download $\binom{N-1}{t-1}$ bits from the desired message $A$, where each bit is from one of the $\binom{N-1}{t-1}$ sub-messages of message $A$. In order to main privacy, we introduce message symmetry, and perform the same downloading operation for all the remaining $(K-1)$ messages. Hence, from each DB, we download a total of $\binom{K}{1}\binom{N-1}{t-1}$ bits, out of which the number of desired bits are $\binom{N-1}{t-1}$. This is also shown in the first row of Table \ref{table4}. 

\vspace{8pt}
\noindent \textbf{Stage $2$}: In the second stage, we download pairs of bits. To this end, let us focus on $DB_1$. From $DB_1$'s perspective, we can download the desired bits of message $A$ along with undesired bits of the remaining $(K-1)$ messages that have been downloaded from the remaining $(N-1)$ databases and are also stored at $DB_1$. We now carefully go over this sequence of steps: a) first note that the number of pairings of a desired message with undesired message is $\binom{K-1}{1}$; b) second, the number of other databases that can be paired with $DB_1$ are $(N-1)$; c) for a fixed pairing with other database (say $DB_i$, $i\neq 1$), the number of sub-messages which are stored at both $DB_1$ and $DB_i$ are $\binom{N-2}{t-2}$. For each such sub-message, the number of undesired bits that were stored at $DB_1$ and were downloaded from $DB_i$ in Stage $1$ are $(t-1)^{0}=1$.  Hence, from $DB_1$ the number of desired bits downloaded in Stage $2$ are $\binom{K-1}{1}(N-1)\binom{N-2}{t-2}(t-1)^{0}$, whereas the total number of downloaded bits are  $\binom{K}{2}(N-1)\binom{N-2}{t-2}(t-1)^{0}$.

\begin{table*}[t]
	\renewcommand{\arraystretch} {1.3}
	\caption{General Storage Constrained PIR Scheme: Total vs. Desired Downloaded Bits (per DB)}\vspace{5pt}
	\label{table4}
	\centering 
	\begin{tabular}{|c|c|c|c|}
		\hline
		Stages & Tuple&Total (Per DB)&Useful (Per DB)\\
		\hline
		Stage $1$ & Single&$\binom{K}{1}\binom{N-1}{t-1}$&$\binom{N-1}{t-1}$\\
		\hline
		Stage $2$ & Pair&$\binom{K}{2}(N-1)\binom{N-2}{t-2}(t-1)^0$&$\binom{K-1}{1}(N-1)\binom{N-2}{t-2}(t-1)^0$\\
		\hline
		Stage $3$ & Triple&$\binom{K}{3}(N-1)\binom{N-2}{t-2}(t-1)^1$&$\binom{K-1}{2}(N-1)\binom{N-2}{t-2}(t-1)^1$\\
		\hline
		.&.&.&.\\
		.&.&.&.\\
		\hline
		Stage $i$ &$i$-tuple&$\binom{K}{i}(N-1)\binom{N-2}{t-2}(t-1)^{i-2}$&$\binom{K-1}{i-1}(N-1)\binom{N-2}{t-2}(t-1)^{i-2}$\\
		\hline
		.&.&.&.\\
		.&.&.&.\\
		\hline
		Stage $K$ &$K$-tuple&$\binom{K}{K}(N-1)\binom{N-2}{t-2}(t-1)^{K-2}$&$\binom{K-1}{K-1}(N-1)\binom{N-2}{t-2}(t-1)^{K-2}$\\
		\hline
	\end{tabular}{}
\end{table*}
\vspace{8pt}
\noindent \textbf{Stage $i$}: We continue to proceed and in the general stage $i$, 
 we download tuples of bits composed of $i$ different messages. Again, focusing from $DB_1$'s perspective, we can download the desired bits of message $A$ along with undesired bits of the remaining $(K-1)$ messages that have been downloaded from the remaining $(N-1)$ databases and are also stored at $DB_1$. We now carefully go over this sequence of steps: a) the number of pairings of a desired message with $(i-1)$ undesired messages is $\binom{K-1}{i-1}$; b) second, the number of other databases that can be paired with $DB_1$ are $(N-1)$; c) for a fixed pairing with other database (say $DB_\ell$, $\ell\neq 1$), the number of sub-messages which are stored at both $DB_1$ and $DB_\ell$ are $\binom{N-2}{t-2}$. For each such sub-message, the number of undesired bits that were stored at $DB_1$ and were downloaded from $DB_\ell$ in Stage $(i-1)$ are $(t-1)^{i-2}$.  Hence, from $DB_1$ the number of desired bits downloaded in Stage $i$ are $\binom{K-1}{i-1}(N-1)\binom{N-2}{t-2}(t-1)^{i-2}$, whereas the total number of downloaded bits are  $\binom{K}{i}(N-1)\binom{N-2}{t-2}(t-1)^{i-2}$.
We similarly continue till $K$ stages. 
Let us now calculate the total number of desired and downloaded bits:
\begin{align}
\text{Desired bits (per DB)}  &=\binom{N-1}{t-1}+\sum_{i=2}^{K}\binom{K-1}{i-1}(N-1)\binom{N-2}{t-2}(t-1)^{i-2}\nonumber\\
&=\binom{N-1}{t-1}+(N-1)\binom{N-2}{t-2}\sum_{i=2}^{K}\binom{K-1}{i-1}(t-1)^{i-2}\nonumber\\
&=\binom{N-1}{t-1}+\frac{(N-1)}{(t-1)}\binom{N-2}{t-2}\sum_{i=2}^{K}\binom{K-1}{i-1}(t-1)^{i-1}\nonumber\\
&=\binom{N-1}{t-1}+\frac{(N-1)}{(t-1)}\binom{N-2}{t-2}(t^{K-1}-1)\nonumber\\
&=\binom{N-1}{t-1}+\frac{(N-1)}{(t-1)}\binom{N-2}{t-2}(t^{K-1}-1)\nonumber\\
&=\binom{N-1}{t-1}+\binom{N-1}{t-1}(t^{K-1}-1)= \binom{N-1}{t-1}t^{K-1} =  \frac{1}{N}\times\binom{N}{t}t^{K}
\end{align}
Since the above is the number is the desired number of bits (per-DB),  hence,  from all the $N$ databases, the user is able to recover all $L=\binom{N}{t}t^{K}$ bits of the desired message. 
\begin{align}
\text{Total downloaded bits (per DB)} &=\binom{N-1}{t-1}\binom{K}{1}+\sum_{i=2}^{K}\binom{K}{i}(N-1)\binom{N-2}{t-2}(t-1)^{i-2}\nonumber\\
&=\binom{N-1}{t-1}\binom{K}{1}+(N-1)\binom{N-2}{t-2}\sum_{i=2}^{K}\binom{K}{i}(t-1)^{i-2}\nonumber\\
&=\binom{N-1}{t-1}\binom{K}{1}+\frac{(N-1)}{(t-1)}\binom{N-2}{t-2}\sum_{i=2}^{K}\binom{K}{i}(t-1)^{i-1}\nonumber\\
&=\binom{N-1}{t-1}\binom{K}{1}+\binom{N-1}{t-1}\sum_{i=2}^{K}\binom{K}{i}(t-1)^{i-1}\nonumber\\
&=\binom{N-1}{t-1}\left(K +\sum_{i=2}^{K}\binom{K}{i}(t-1)^{i-1}\right)\nonumber\\
&=\binom{N-1}{t-1}\left(K +\frac{1}{t-1}\sum_{i=2}^{K}\binom{K}{i}(t-1)^{i}\right)\nonumber\\
&=\binom{N-1}{t-1}\left(K +\frac{1}{t-1}(t^{K}- 1 - (t-1)K)\right)\nonumber\\
&=\binom{N-1}{t-1}\frac{t^{K}-1}{t-1}=\binom{N-1}{t-1}(1+ t+ t^{2}+\ldots + t^{K-1})\nonumber
\end{align}
Hence, the download cost $D(\mu)$ of the storage constrained PIR scheme when $\mu=t/N$ is given as
\begin{align}
D(\mu)&= \frac{N\times \text{Total Downloaded bits (per DB)}}{N \times \text{Desired bits (per DB)}}\nonumber\\
&= \frac{\binom{N-1}{t-1}(1+ t+ t^{2}+\ldots + t^{K-1})}{\binom{N-1}{t-1}t^{K-1}}= 1 + \frac{1}{t} + \frac{1}{t^{2}}+ \ldots + \frac{1}{t^{K-1}}.
\end{align}
This completes the proof of Theorem \ref{theorem1}.
\vspace{-15pt}
\section{Conclusion}
\vspace{-10pt}
In this work, we investigated the PIR problem from storage constrained databases. The main contribution of this work is a general achievable scheme for PIR from storage constrained databases for any value of storage. In particular, for any $(N,K)$, with normalized storage $\mu= t/N$, where the parameter $t$ can take integer values $t \in \{1, 2, \ldots, N\}$, we show that our proposed PIR scheme achieves a download cost of $\left(1+ \frac{1}{t}+ \frac{1}{t^{2}}+ \cdots + \frac{1}{t^{K-1}}\right)$. Furthermore, the lower convex hull of all such $(D(\mu), \mu)$ pairs is also achievable and the scheme strictly outperforms memory sharing between the extreme values of $\mu$. There are several interesting directions for future work on this important variation of PIR. An immediate interesting direction would be to obtain lower bounds on the PIR capacity (or download cost) as a function of the storage.  Moreover, it would be interesting to investigate storage-constrained PIR problems by considering scenarios such as (i) colluding databases; and (ii) having additional constraints on storage (such as data recoverability constraints from any $\ell$ out of $N$ databases). 

\vspace{-15pt}

\bibliographystyle{unsrt}
\bibliography{paper}

\end{document}